\documentclass[prl,aps,amsmath,amssymb,twocolumn,groupedaddress,floats,showpacs,final,reprint]{revtex4-1}
\usepackage{graphicx}
\usepackage{amssymb,amsmath}
\usepackage{color}
\usepackage{bm}
\begin{document}

\author{Janik Sch{\"o}nmeier-Kromer}
\affiliation{Department of Physics, Arnold Sommerfeld Center for Theoretical Physics and Center for NanoScience, University of Munich, Theresienstrasse 37, 80333 Munich, Germany}
\author{Lode Pollet}
\affiliation{Department of Physics, Arnold Sommerfeld Center for Theoretical Physics and Center for NanoScience, University of Munich, Theresienstrasse 37, 80333 Munich, Germany}


\title{Ground state phase diagram of the 2d Bose-Hubbard model with anisotropic hopping}


\date{\today}
\begin{abstract}
We compute the ground state phase diagram of the 2d Bose-Hubbard model with anisotropic hopping using quantum Monte Carlo simulations, connecting the 1d to the 2d system. We find that the tip of the lobe lies on a curve controlled by the 1d limit over the full anisotropy range while the universality class is always the same as in the isotropic 2d system. This behavior can be derived analytically from the lowest RG equations and has a shape typical for the underlying Kosterlitz-Thouless transition in 1d. We also compute the phase boundary of the Mott lobe at unit density for strong anisotropy and compare it to the 1d system. Our calculations shed light on recent cold gas experiments monitoring the dynamics of an expanding cloud.
\end{abstract}



\maketitle


In one dimension, interactions and quantum fluctuations are stronger than in any other dimension. As a consequence, single-particle excitations cannot occur and only collective excitations are formed. The liquids in 1d are known as Luttinger liquids~\cite{Haldane}, which are critical phases with algebraic correlations. They are for instance susceptible to a lattice when they can form a Mott insulator or to disorder when they can form a Bose glass~\cite{GS}. A particular kind of relevant perturbations are the ones leading to a dimensional crossover in which Luttinger liquids or 1d systems are coupled with each other in a matrix of higher dimension~\cite{Giamarchi2010}, which can be experimentally realized~\cite{Stoferle, Schori}. We will consider hopping processes (that is, a Josephson coupling) between tubes of scalar bosonic Luttinger liquids and Mott insulators arranged in a 2d setup, where the intertube coupling is varied from zero to an equally strong value as the intra-tube coupling. For this 1d-2d crossover, mean-field theory fails~\cite{Ho2004, Cazalilla2006, Gangardt2006}, and this topic is one of the remaining open problems in the 1d world~\cite{Cazalilla2011, Giamarchi2010}.

Recently, studies on dimensional crossovers for fermions have come to the front of attention again. The Mott transition in a frustrated Hubbard model with next-nearest neighbor hopping at half filling on a quasi 1d lattice was studied in Ref.~\cite{Marcin2012}, featuring a closing of the 1d Mott gap and supporting the idea that superconductivity is mediated by magnetic fluctuations in organic salts~\cite{Jerome}. The situation is reminiscent of the pseudogap phase in high-$Tc$ cuprates where enhanced spin fluctuations and spatial correlations in the copper oxide planes occur in the proximity of an insulating phase. The study was extended to dynamical correlation functions in Ref.~\cite{Marcin2013} providing evidence for a dimensional-crossover-driven confinement of spinons~\cite{Lake2010}. A  3d Hubbard model with anisotropic hopping was studied with cluster dynamical mean field methods and yielded results in agreement with a cold gas realization for temperatures down to the hopping amplitude~\cite{Imriska2013}.

In a recent cold gas experiment by Ronzheimer {\it et al.}~\cite{Ronzheimer2012} a bosonic system was prepared in the atomic limit in 2d with precisely one particle per site. The experimentalists then quenched the interaction and the hopping along the $x$ and $y$ directions to their respective target values, switched off the confinement, and monitored the dynamics of the density. In the pure 1d case the dynamics is ballistic for free bosons and for hard-core bosons (which can be mapped to free fermions). For the isotropic 2d case, a system quenched to parameters that remain in the Mott insulating regime showed a strongly suppressed expansion. For the anisotropic case, the asymptotic velocity of the width of the density cloud (the core velocity) shows first a decrease, then an upturn and finally a plateau as a function of interaction strength.
For experiments like this one, as well as for the study of Ref.~\cite{Vidmar2013} it would hence be very useful to know the thermodynamic phase diagrams in the 1d-2d crossover regime.

In this Letter, we address the above questions and provide an analytical estimate for the behavior of the tip of the Mott lobe as a function of anisotropy based on a RG treatment around the 1d limit, supported by large-scale Monte Carlo simulations. 
We also compute the boundaries of the Mott lobe at unit density for strong anisotropy, which nevertheless shows strong signatures of the 2d case. These findings have profound implications on present cold gas experiments as well as for any condensed matter system in which there is a strong coupling in one dimension and only a weak Josephson coupling in other directions.

We consider the Bose-Hubbard model on a square lattice with anisotropy in the hopping,
\begin{eqnarray}
H - \mu N &= &  \sum_{\bm{r}=(i,j)} -t_x (b^{\dagger}_{i,j} b_{i+1,j} + {\rm hc}) - t_y  (b^{\dagger}_{i,j} b_{i,j+1} + {\rm hc}) \nonumber \\
{}  & {} &   + \frac{U}{2} \sum_{\bm{r}} n_{\bm{r}} (n_{\bm{r}} - 1) - \mu \sum_{\bm{r}} n_{\bm{r}}. \label{eq:ham}
\end{eqnarray}
Lattice coordinates are denoted by 2-tuples $\bm{r} = (i,j)$, the hopping amplitudes in the $x$ and $y$ direction are $t_x$ and $t_y$, respectively, the on-site interaction of density-density type has a strength $U$ and the chemical potential is $\mu$. 'hc' denotes the Hermitean conjugate. The unit is $t_x = 1$ unless written otherwise. When $t_y = 0$, the system is purely 1d, when $t_y = 1$, the system is an isotropic 2d system. We will vary $t_y$ between 0 and 1 in this study, and compute the zero temperature phase diagram.  The Bose-Hubbard model is the simplest bosonic system featuring a transition between a superfluid (SF) and a Mott insulating (MI) phase~\cite{Fisher89, Sachdev2011}. There are two different types of transitions: First, away from the tip of the lobe, the transition is driven by density. Its universality class is the same as that of the dilute Bose gas and has a dynamical exponent $z=2$. Second, at the tip of the lobe a trajectory of constant density can be followed, in which case the transition is purely interaction driven. It belongs then to the universality class of the $(d+z)$ dimensional XY-model, with dynamical exponent $z=1$. In case of anisotropy, the nature of the phase transitions is not expected to change. 
This system has been studied in Ref.~\cite{Bergkvist2007, Rehn2008} but only for a system of finite length and thus far away from the thermodynamic limit. Also a strong coupling expansion has been undertaken~\cite{Freericks2008}.


The mean-field decoupling approximation~\cite{Fisher89, Sachdev2011} predicts that the tip of the lobe follows a linear behavior. However, the phase diagram in the 1d case shows reentrant behavior near the tip of the lobe caused by the cusp at the tip, located at $U_c^{1d} = 3.25(5) t_x$, which is typical of the Kosterlitz-Thouless transition and not captured by mean-field theory~\cite{Kuhner1998}. Deviations from the linear behavior must hence occur at least in the limit $t_y \to 0$.
We address therefore the 1d-2d crossover theoretically starting from the 1d limit. Using a bosonization and a RG approach for the 1d-3d crossover, Ref.~\cite{Ho2004, Cazalilla2006} found an analytical expression, namely a power-law behavior, although they had to neglect the renormalization of the Luttinger parameter $K$   (Eq.~\ref{eq:renorm_K}). As already noticed in Ref.~\cite{Ho2004, Cazalilla2006}, this approximation is too crude for us because $K$ changes rapidly near the critical point. A numerical evaluation of the full RG equations was hence provided, with a substantially different answer. The flow equations in dimensionless form to lowest order~\cite{Ho2004, Cazalilla2006} are
\begin{eqnarray}
\frac{dK}{d\ell} & = &  -g_u^2 K^2 + 2 g_J^2 \label{eq:renorm_K} \\
\frac{dg_u}{d\ell} & = & (2 - K)g_u  \label{eq:renorm_u} \\
\frac{dg_J}{d\ell} & = & (2 - 1/2K) g_J. \label{eq:renorm_j} 
\end{eqnarray}
Here, $g_u$ and $g_J$ are the  interaction strength and Josephson coupling, respectively. 
The flow parameter is $\ell = \ln R$, with $R$ the characteristic length scale in the quasi-1d limit, 
and $K(\ell)$ is the mesoscopic value of the Luttinger parameter given by $K = \pi \sqrt{ \rho_s \kappa}$, where $\rho_s$ is the superfluid density and $\kappa$ the compressibility. The Eqs.~\ref{eq:renorm_u} and \ref{eq:renorm_j} follow from the scaling dimension of these operators. Their left hand sides are proportional to  the number of vortex pairs of size $\sim R$ in an area $\sim R^2$ in the quasi-1d limit. 
The induced bosonic interaction between the chains is absent in the pure 1d system, and remains small when the 1d tubes are coupled~\cite{Cazalilla2006}. It can be neglected.

By viewing this set of flow equations as the one coming from a double sine-Gordon model and realizing the different length scales at which $g_J$ and $g_u$ may become relevant, an analytic expression~\cite{thanks} for the behavior of the tip of the lobe can be found along the same lines as the analysis of the 1d disordered Bose-Hubbard model at commensurate densities of Ref.~\cite{Svistunov96} and partly Ref.~\cite{Kashurnikov96}. The starting point is that we couple 1d Mott insulators with Josephson junctions at a mesoscopic scale where $K(\ell) \approx 2$. The coupling will effectively modify $K(\ell)$ starting at some cross-over scale $\ell_*$. Note that the Josephson coupling $g_J$ becomes relevant for $K = 1/4$. In the initial stages when $\ell \ll \ell_*$, we can put $g_J$ to zero in the RG equations and use the solution for $K$ on the insulating side of the 1d side (cf. Ref.~\cite{Svistunov96}),
\begin{equation}
K^{-1}(\ell) = \frac{1}{2} + a \tan[4 a \ell - \pi/2].
\label{eq:Lutt_solution_MI}
\end{equation}
Here, $a$ depends on the system parameters as $a \propto \sqrt{U / U_c^{1d} - 1}$. We see that $K(\ell)$ keeps its value close to 2 until $\ell$ comes close to $\pi/4a$, when $K^{-1}$ diverges. Thus, $\ell_* =\pi/(4a)$ determines the 1d-2d crossover scale. Application of Eq.~\ref{eq:Lutt_solution_MI} at $K=1/4$ is justified by the rapid change in behavior of $K^{-1}(\ell)$ in the vicinity of $\ell_*$.
This leads to the following estimate for the SF-MI transition line,
\begin{equation}
t_y^c/t_x(U) \propto \exp \left( - \frac{\pi s}{4 b \sqrt{U / U_c^{1d} - 1}}  \right),
\label{eq:transition_line}
\end{equation}
with $b$ a constant and $s$ the exponent which expresses how the dimensionless coupling scales with the characteristic length in dimensionless units. In 2d, superfluidity is destroyed by vortex loops scaling as $\sim R_*^2$ and thus $s=2$.


To illustrate the theory, we study the model Eq.~\ref{eq:ham} numerically by path integral Monte Carlo simulations with worm-type updates~\cite{Prokofev1998} in the implementation of Ref.~\cite{Pollet2007} providing a statistically exact answer. For a recent overview of the method with applications to cold gas systems, see Ref.~\cite{Pollet2012_review}. These methods allow the computation of the superfluid density via the winding number fluctuations~\cite{Pollock1987}, which distinguishes between the SF and the MI. In case of anisotropic hopping, the winding number fluctuations are also anisotoropic, $ \langle W_x^2 \rangle \neq \langle W_y^2 \rangle$. Correspondingly, the system size anisotropy was chosen as
\begin{equation}
\frac{L_y}{L_x} = \sqrt { \frac{t_y}{t_x}},
\label{eq:scaling_L}
\end{equation}
such that the respective helicity moduluses are about the same in magnitude and the winding number fluctuations scale similarly along $x$ and $y$ directions~\cite{Wang2012}. In order to determine the tip of the lobe, we only performed Monte Carlo measurements when the density was commensurate. 
Since the winding number is an integer it is scale invariant at the transition point, and simulations performed for different system sizes should provide curves for the winding number fluctuations that cross in one point.

\begin{figure}[tpb]
\includegraphics[angle=-90, width=1.0\columnwidth]{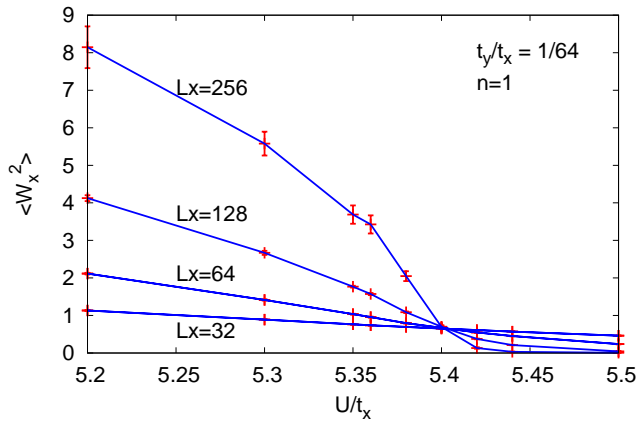}
\caption{\label{fig:FSS_tip}
(Color online). Finite size scaling of the winding number fluctuations along $x$ for $t_y/t_x = 1/64$ with Monte Carlo measurements in the canonical ensemble at $n=1$. System scaling parameters are $\beta/L_x = 1/2$ and $L_y / L_x = 1/8$, and the system size in the $x$-direction is mentioned in the figure. \label{fig_FSS_tip}
}
\end{figure}

This can be seen in Fig.~\ref{fig_FSS_tip} where the position of the tip of the lobe has been determined for $t_y / t_x = 1/64$ resulting in a critical value $U_c/t_x = 5.404(2)$. This anisotropy is a typical value for what is experimentally claimed to be a 1d system as in {\it e.g.} Ref.~\cite{Ronzheimer2012}, but this means that the location of the tip of the lobe would be off by $\sim 60\%$ compared to the true 1d value in a hypothetical experiment with these parameters! The slope of the curves increases in a way compatible with the finite size scaling predictions, showing that all system sizes are in the scaling regime. It is worth noting that the curves for the winding number squared along the $y$-direction cross at the same point within error bars. This is expected on the basis of the arguments presented above, and rules out an elusive sliding phase (which is a Luttinger liquid in the axial direction and a Mott insulator along the transverse direction~\cite{Emery2000}) or any other exotic phase ({\it e.g.}, a supersolid) in the thermodynamic limit~\cite{Cazalilla2006}.

\begin{figure}[tpb]
\includegraphics[angle=-90, width=1.0\columnwidth]{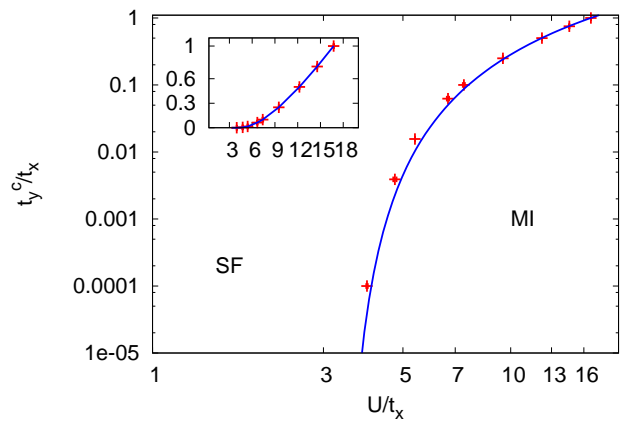}
\caption{\label{fig:crossing}
(Color online).  Critical value of the hopping amplitude along $y$ as a function of interaction strength $U$ for the tip of the $n=1$ Mott lobe on a linear-linear plot in the inset and on a log-log plot in the main figure. Error bars are shown but are barely visible. The full line shows the quality of the analytical prediction, Eq.~\ref{eq:transition_line},  where the overall prefactor  and $b = 0.141(4)$ are the fit parameters.
\label{fig_tip}
}
\end{figure}

The curve describing the location of the tip of the MI lobe as a function of anisotropy is shown in Fig.~\ref{fig:crossing}. 
For the strongest anisotropy, $t_y / t_x = 10^{-4}$, system sizes $L_x = 400, 600$ and $800$ were used. The larger error bars (but barely visible on the scale of the plot) reflect a systematic uncertainty because of the numerical difficulties inherent to the closeness of the 1d system. To test the analytical arguments, we plot the location of the tips on the log-log scale in the main part of the figure along with the prediction Eq.~\ref{eq:transition_line}, in which only the overall prefactor and the parameter $b$ are fitting parameters. Remarkably, the fit is very good not only at low values of $t_y$ but extends all the way to $t_y = 1$. 
Because $g_J$ is a relevant perturbation, the values of $t_y^c$ are exponentially smaller than the gaps $\Delta_g(U)$ of the pure 1d system in the MI phase, which is also of the form Eq.~\ref{eq:transition_line} but with $s=1$ originating form the linear length scale for the characteristic vortex-antivortex pair separation~\cite{Fisher89, Svistunov96}.

\begin{figure}[tpb]
\includegraphics[angle=-90, width=1.0\columnwidth]{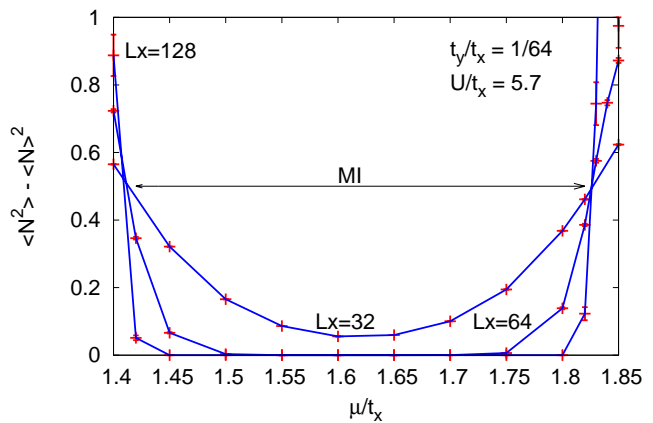}
\caption{\label{fig:FSS_lobe}
(Color online).  Finite size scaling of the winding number fluctations in the $\tau$-direction (equivalent to particle number fluctuations and thus the compressiblility when divided by temperature) for $t_y / t_x = 1.64$ and $U/t_x = 5.7$. Parameters are $\beta = 16$ for $L_x = 32$, $\beta = 64$ for $L_x = 64$ and $\beta = 256$ for $L_x = 128$. The ratio $L_y/L_x = 1/8$ was kept fixed.
\label{fig:compr}
}
\end{figure}

\begin{figure}[tpb]
\includegraphics[angle=-90, width=1.0\columnwidth]{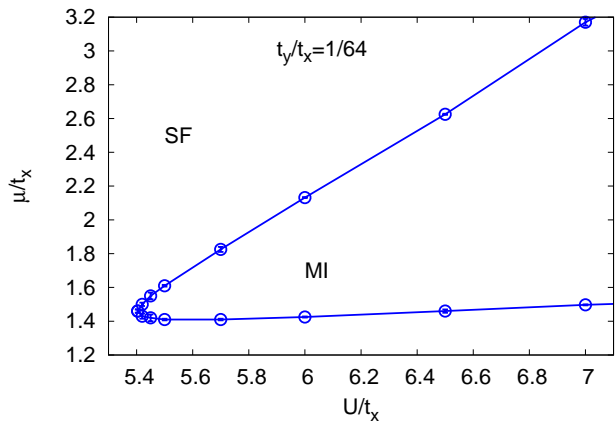}
\caption{\label{fig:Mottlobe}
(Color online).  The $n=1$ Mott lobe is shown in the $(U, \mu)$ plane for $t_y / t_x = 1/64$ in the vicinity of the tip.
\label{fig:MI_lobe}
}
\end{figure}

We also determined the shape of the MI lobe. To this end, we worked in the grand-canonical ensemble and used fluctuations in the total particle number ($\Delta N = \langle N^2 \rangle - \langle N \rangle^2$) to distinguish between the phases. This is, up to a factor of the inverse temperature $\beta = 1/T$, related to the compressibility, which is zero in the MI and finite in the SF. They can be thought of as winding number fluctuations in the imaginary time direction. Away from the tip of the lobe, we scale $\beta \to 4 \beta$ when $L_x \to 2L_x$. To start this flow we take the lowest $\beta$ to be half of the smallest system size. 
The quantity $\Delta N$ is  scale invariant at the transition point. The finite size scaling close to the tip of the lobe for a relatively strong anisotropy $t_y / t_x$ is shown in Fig.~\ref{fig:compr}. For values of $U$ very close to the tip of the lobe, the finite size analysis becomes more cumbersome due to the vicinity of the tip which scales with $z=1$. In particular, for $U/t_x = 5.42$ and $U=5.45$ a precise determination of the critical point is more difficult and has hence slightly larger relative error bars than for larger values of $U/t_x$. We are limited to  a maximum system size $L_x = 192, L_y = 24$ and $\beta = 576$ due to computer technical reasons. Interestingly, curves for $L_x = 64$, $L_x = 128$ and $L_x = 192$ have a strong overlap in the SF phase on approach to the critical point before dropping off, reminiscent of the nearby 1d Kosterlitz-Thouless physics.

The boundary of the Mott lobe for $t_y / t_x = 1/64$ is shown in Fig.~\ref{fig:MI_lobe}. Clearly, the shape near the tip of the lobe is round. Gaps in the MI can be determined from the distance between the 2 phase transition lines. Also when the phase diagram is plotted in the $(t_x/U, \mu/U)$ plane, no signs of reentrant behavior are seen.

The phase diagrams presented in this work can straightforwardly be obtained in a cold gas experiment, provided the temperature is low enough. In a time-of-flight interference experiment, the superfluid phase will have interference peaks at zero momentum and for all reciprocal lattice vectors. The system sizes are best taken according to Eq.~\ref{eq:scaling_L}.  With a single-site precision in-situ measuring microscope~\cite{Greiner1, Greiner2, Kuhr} the density and the compressibility could be measured. It is thus possible to obtain the phase diagrams in a cold gas experiment with existing technology. 
Regarding the expansion experiment of Ref.~\cite{Ronzheimer2012} the location of the tip of the lobe as a function of anisotropy correlates qualitatively well (except for the strongest anisotropy) with the point where the plateaus for large $U$ in Fig. 3 of Ref.~\cite{Ronzheimer2012} begin, {\it i.e.}, where the expansion speed is insensitive to an increase in $U$. The quench energy is then smaller than the gap, and the core velocity should be identical when $U$ increases further. The other characteristic scale in Fig. 3 of Ref.~\cite{Ronzheimer2012} is the upturn in the core velocity for all strong anisotropies around $U/t_x \approx 3$, which coincides with the scale where the 1d system goes over in the MI phase. This upturn is markedly different from the behavior of the isotropic 2d liquid. However, on the time scale of the experiment the correlations in the $y$ direction have not fully developed. In 1d, the time scale on which the asymptotic behavior sets in is approximately $20/t_x$~\cite{Ronzheimer2012}; assuming this holds for the 1d-2d crossover as well and taking the experimental observation time window of about $ 60 /t_x$ into account, it is clear that correlations along $y$ cannot have developed unless for the isotropic case and $t_y / t_x = 0.5$. For stronger anisotropy, the system's dynamics will thus be strongly influenced by the effective 1d character on the accessible time scales.
Additional experimental data (with {\it e.g.} a hold time after the interaction quench before opening the trap such that correlations can develop better), would be needed to describe this crossover in more detail, or indentify a sliding phase valid only at mesoscopic scales.


In conclusion, we have studied the zero temperature phase diagram of the 1d-2d crossover of the scalar Bose-Hubbard model. Although the universality class at the tip of the Mott insulator lobe is always the same as for the isotropic 2d model, the location of the quantum critical point follows the behavior given by 1d physics over the entire anisotropy range: the RG flow equations are those of a double sine-Gordon model operating at different length scales that can consequently be analyzed independently. These analytic expressions likewise determine the scale at which a mesoscopic system can still be considered 1d.
We also computed the boundary of the Mott insulator and found no signs of reentrant behavior for an anisotropy down to $t_y / t_x = 1/64$. 
Our calculations can directly be verified in a cold gas experiment, and provide new insight in recent cold gas experiments exploring dynamics in the crossover regime~\cite{Ronzheimer2012, Vidmar2013}. In future work, the present analysis can be extended to finite temperatures and the 1d-3d crossover.  The detection of a Higgs amplitude mode, which became of recent experimental interest~\cite{Higgs1, Higgs2}, may be advantageous in the presence of anisotropy, as has been suggested in Ref.~\cite{Cazalilla2006}. Such an analysis would require a straightforward extension of Ref.~\cite{Pollet2012_Higgs} in combination with this work.

Acknowledgements -- We are grateful to Thierry Giamarchi, Fabian Heidrich-Meisner, Ulrich Schneider, Boris Svistunov and Lev Vidmar for inspiring discussions. 
This work is supported by the Excellence Cluster NIM, FP7/Marie-Curie Grant No. 321918 ("FDIAGMC"),  FP7/ERC Starting Grant No. 306897 ("QUSIMGAS") and by a grant from the Army Research Office with funding from DARPA. Use was made of the ALPS libraries for error evaluation~\cite{ALPS2}.

\end{document}